\documentclass[aps,pra,superscriptaddress,amsmath,amssymb,twocolumn,floatfix]{revtex4-2}

\usepackage{xcolor}
\usepackage{graphicx}
\usepackage{braket}
\usepackage{float}
\usepackage{units} 
\usepackage{soul} 
\usepackage{verbatim} 	
\usepackage[hidelinks]{hyperref}
\usepackage{bm}
\usepackage{bigints}
\usepackage{tikz,pgfplots}
\usepackage{amsfonts}
\usepackage{amssymb}
\usepackage{upgreek}
\usepackage[T1]{fontenc}

\pgfplotsset{compat=1.18}
\begin{document}

\title{Shaping Cold Atom Clouds with a Vortex Beam}

\author{Arianna Bertoluzza}
\email{arianna.bertoluzza@uni-tuebingen.de}
\affiliation{Center for Quantum Science, Eberhard Karls Universit\"at T\"ubingen, Auf der Morgenstelle 14, 72076 T\"ubingen, Germany}
\author{Sonja Lorenz}
\affiliation{Center for Quantum Science, Eberhard Karls Universit\"at T\"ubingen, Auf der Morgenstelle 14, 72076 T\"ubingen, Germany}
\author{Paul Hampp}
\affiliation{Center for Quantum Science, Eberhard Karls Universit\"at T\"ubingen, Auf der Morgenstelle 14, 72076 T\"ubingen, Germany}
\author{Moriz H\"arle}
\affiliation{Center for Quantum Science, Eberhard Karls Universit\"at T\"ubingen, Auf der Morgenstelle 14, 72076 T\"ubingen, Germany}
\author{Daniel Braun}
\affiliation{Center for Quantum Science, Eberhard Karls Universit\"at T\"ubingen, Auf der Morgenstelle 14, 72076 T\"ubingen, Germany}
\author{David Petrosyan}
\affiliation{Center for Quantum Science, Eberhard Karls Universit\"at T\"ubingen, Auf der Morgenstelle 14, 72076 T\"ubingen, Germany}
\affiliation{Institute of Electronic Structure \& Laser, FORTH, Heraklion 70013, Crete, Greece}
\author{J\'{o}zsef Fort\'{a}gh}
\affiliation{Center for Quantum Science, Eberhard Karls Universit\"at T\"ubingen, Auf der Morgenstelle 14, 72076 T\"ubingen, Germany}
\author{Andreas G\"unther}
\email{a.guenther@uni-tuebingen.de}
\affiliation{Center for Quantum Science, Eberhard Karls Universit\"at T\"ubingen, Auf der Morgenstelle 14, 72076 T\"ubingen, Germany}

\begin{abstract}
\noindent
We introduce a method for shaping a cold atom cloud using a vortex laser beam with a polarization singularity at its center, which creates a point of vanishing intensity. Exploiting this feature we experimentally demonstrate two different schemes to create micron-scaled line- and sheet-like atomic density distributions. In the \textit{dynamic scheme}, atoms in the bright beam regions are accelerated and therefore effectively removed from the cloud. In the \textit{dark-state scheme}, these atoms are pumped into a state that does not interact with the shaping light. In both cases, an atomic distribution remains, either as a thin line or as a sheet when an additional polarizer is used. We find good agreement between the experimental results and our theoretical model, which predicts the method to be in principle not diffraction-limited, paving the way for studies of phenomena arising in unconfined atomic ensembles on the micrometer scale.
\end{abstract} 

\date{\today}

\maketitle

\noindent

\section{Introduction}
Over the last decade, the shaping of cold atom clouds on the micron-scale has become a powerful tool for studying interacting quantum systems \cite{gross2017quantum,browaeys2020,henriet2020,kaufman2021}. To this end, atomic ensembles are typically confined in magnetic or optical traps, which can be engineered to almost arbitrary spatial scales. While macroscopic magnetic confinement can be achieved via conventional coil systems \cite{metcalf1994}, the micron-sized regime is typically accessed in chip-based structures \cite{folman2002,fortagh2007,reichel2011}, with a lower limit in the structure size being given by current constrains and the distance of the cloud to the field generating elements. Optical confinement, on the other hand, allows for a large variety of trapping schemes, ranging from simple single-beam and cross-beam dipole traps \cite{grimm2000} to more complex structures, as created in optical lattices \cite{jessen1996optical} or tweezer arrays \cite{muldoon2012}. Thin atomic distributions and periodic arrangements can then be routinely created \cite{endres2016}, laying the foundation for groundbreaking experiments such as studies of Anderson localization \cite{Roati2008,Billy2008}, topological states \cite{cooper2019}, or the Mott insulator transition \cite{greiner2002}.
While such confining potentials have been extensively studied, the shaping of unconfined atom clouds on the micrometer scale has not yet attracted as much attention, despite the promising prospects offered by structured illumination of atomic ensembles.
On the one hand, the preparation and purification of atomic gases during MOT loading has been successfully demonstrated in the past by combining structured illumination with optical pumping and scattering forces, as in the case of the dark SPOT \cite{Ketterle2,Radwell} and the core-shell MOT \cite{Lee}.
On the other hand, techniques such as stimulated emission depletion microscopy \cite{Hell} and saturated structured illumination microscopy \cite{Mats} have been shown to provide resolution capabilities beyond the diffraction limit. This has motivated studies with structured illumination of ultracold gases, for example in the areas of imaging \cite{McDonald} and creation of optical lattices \cite{Wang}, achieving enhanced resolution compared to the diffraction limit.

Here, we introduce an approach for creating thin, micrometer-scale line- and sheet-like atomic ensembles which combines the high resolution capabilities of structured illumination with the state-selective nature of optical pumping and scattering forces. Our work considerably differs from previous studies, since atoms in the shaped cloud are both unconfined and disordered.
In particular, our technique is based on shaping an atomic cloud via a vector vortex beam with a polarization singularity at the beam center, where the beam intensity vanishes. This allows for the selective interaction between the light field and the atoms in the outer beam regions, which can be used to effectively remove (hide) atoms and thus shape the cloud.
At lowest order approximation, the intensity profile close to the vortex beam center is parabolic, with the curvature scaling linearly with the beam power.
Therefore, the size of the remaining atomic distribution can be decreased arbitrarily, even below the diffraction limit, by just increasing the beam power, creating a thin line of atoms.
Additionally, the inhomogeneous polarization distribution of the vortex beam can be used to create more complex spatial intensity distributions, such as a double-lobed structure featuring a zero-intensity line passing through the center. Then, the resulting beam can be used for shaping atomic clouds into sheet-like distributions.

We experimentally demonstrate two different shaping schemes.
In the \textit{dynamic scheme}, atoms in bright beam regions are accelerated due to the scattering force from the resonant shaping beam, selectively removing atoms across the beam profile. In the \textit{dark-state scheme}, the shaping beam frequency is tuned such that atoms in bright beam regions are pumped to a non-interacting dark state, transparent to the shaping beam. 
In both cases, this leads to the formation of either a micron-scaled line of atoms or, when using an additional polarizer, an atomic sheet.
Our experimental results, for a double-lobed beam in the case of the \textit{dynamic scheme} and for a vortex beam in the case of the \textit{dark-state scheme}, show good agreement with the theory, which predicts the method to be in principle not diffraction-limited.
Due to the very simple setup and the low power required for the shaping beam, our method makes the generation of micron-scaled, unconfined clouds easily accessible and provides an experimental platform and alternative approach for studies with thin ensembles of atoms at random positions. This opens the route for the study of phenomena which in the past have only been investigated in ordered and confined systems, such as Rydberg crystallization \cite{Pohl,Schauß} and long-range quantum correlations \cite{Weimer}.

The manuscript is structured as follows. Section \ref{section:theory} provides a theoretical treatment of a cold atom cloud interacting with a vortex beam, including the analytic description of a vector vortex beam with a polarization singularity (\ref{subsection:vortex_beams}), the general idea for cloud shaping (\ref{subsection:cloud_shaping}) and the theoretical description of the \textit{dynamic}- (\ref{subsection:dynamic_shaping_scheme}) and \textit{dark-state shaping scheme} (\ref{subsection:dark_state_pumping_scheme}).
The experimental setup and the shaping sequence are described in Section \ref{section:Experiment}, followed by measurements discussed in Section \ref{sec:results}, where we present a comparison between experiment and theory for both the \textit{dynamic}- (\ref{subsection:dynamic_results}) and the \textit{dark-state scheme} (\ref{subsection:dark_state_results}). In Section \ref{section:conclusion}, we discuss about the fundamental limitations of our shaping method and compare it with diffraction-limited systems.

\section{Theory}
\label{section:theory}

\subsection{Vortex Beams} 
\label{subsection:vortex_beams}

\begin{figure*}[!t]
	\centering
	\includegraphics[]{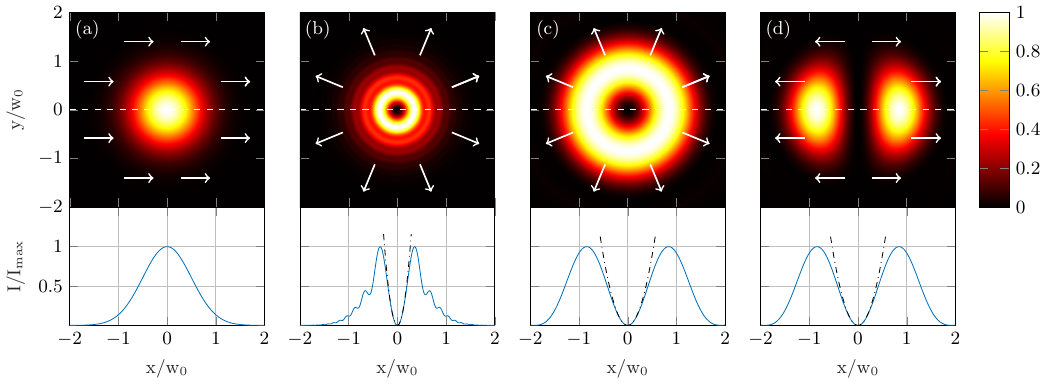}
	\caption{\textbf{Vortex beam.} Intensity profiles of a vortex beam generated from a Gaussian beam (TEM$_{00}$) using a $m=1$ vortex retarder placed at the position of the beam waist ($z'=0$). The profiles are calculated at positions (a) $z=0$, (b) $z=0.05\,z_0$, (c) $z=0.5\,z_0$, and (d) $z=0.5\,z_0$ with additional polarization filter. The 2D profiles are shown together with 1D line-scans through the beam center ($y=0$), with the line-scan positions marked in the 2D profiles (white dashed lines). The line-scans in (b)-(d) are approximated by parabolas around the beam center (black dash-dotted line), with curvatures calculated from Eq.~\eqref{eq:curvatureSimple}. All intensities are normalized to the maximum intensity and all lengths are in units of the initial beam waist $w_0$. The white arrows in the 2D profiles indicate the direction of the local light field polarization.}
	\label{fig:beam_generation}
\end{figure*}

Vortex beams are beams featuring a point of vanishing intensity at the center. This is caused by the presence of either a phase- or a polarization singularity, which originate from a spatially inhomogeneous phase- or polarization distribution with 
cylindrical symmetry, respectively \cite{Ruchi}.
Common methods for creating phase singularities make use of spiral phase plates \cite{Sueda}, computer-generated holograms \cite{Heckenberg}, spatial light modulators \cite{Yang} or optical cavities \cite{Kim}, while the generation of polarization singularities requires the use of vortex retarders \cite{McEldowney}, q-plates \cite{Fu} or multi-beams interference strategies \cite{Mendoza-Hernandez}.
Mathematically, both kinds of vortex beams can be described in general as Laguerre-Gaussian modes LG$_{pl}$, with the radial index $p$ typically considered as $p=0$ and the azimuthal index $|l|\geq1$ indicating the topological charge carried by the singularity \cite{Wang2}. In both cases, for a homogeneous or inhomogeneous polarization pattern, the intensity distributions is the same, but the fieldcan be obtained from the solution of the scalar- or vector Helmholz equation, respectively \cite{Zhan}.
In general, cylindrical vector vortex fields with arbitrary polarization patterns (fully radial, fully azimuthal or mixed) can be described as superposition of scalar LG$_{pl}$ or Hermite-Gaussian (HG$_{mn}$) modes with appropriate polarizations \cite{Maurer}. 

In this work, we employ a vector vortex beam with a polarization singularity produced by a vortex retarder, whose intensity profile resembles in first approximation the one of a LG$_{01}$ mode with a radially symmetric polarization distribution. Mathematically, this can be expressed as a superposition of a scalar HG$_{10}$ and a HG$_{01}$ mode with orthogonal polarizations. However, this description is only valid in the far-field and even there, the overlap with a vortex beam generated by a vortex retarder is incomplete \cite{Beijersbergen}. 
Hence, in order to describe the real intensity profile of our experimental beam, we need to use either numerical or analytical treatments. These allow to derive the intensity at any point along the beam propagation direction starting from the electric field calculated in a single plane, typically taken to be the plane in which the vortex pattern is generated.
Both approaches are presented in detail in Appendix \ref{appendix:VortexAnalytic}, which focuses on the case of a vortex beam generated by a vortex retarder, but whose results can be easily adapted to the case of a phase singularity produced by a spiral phase plate. Here, we report only the final analytical results.

We start by considering a Gaussian TEM$_{00}$ beam with amplitude $E_0$ propagating along the z-axis, with the beam waist $w_0$ at $z=0$ and linear polarization along the x-axis (cf.~Fig.~\ref{fig:beam_generation}a), and a vortex retarder of first order placed in the z$'$-plane and having negligible thickness. In Appendix \ref{appendix:VortexAnalytic}, we show that the electric field at any position after a vortex retarder of first order can be expressed in cylindrical coordinates $\vec{r}\rightarrow (\rho,\varphi,z)$ as
\begin{equation}
\label{eq:EfarfromVR2}
    \Vec{E}(\vec{r}) = \tilde{E}(\rho,z)  \, \begin{pmatrix} \cos{\varphi} \\ \sin{\varphi} \end{pmatrix},
\end{equation}
with
\begin{eqnarray}
    \tilde{E}(\rho,z) = - \sqrt{\pi} E_0 \frac{w_0}{w(z')} \frac{R_c^3(z') k^2}{8(z-z')^2} \, \rho\, e^{-\frac{R_c^2(z') k^2}{8(z-z')^2} \rho^2} e^{\frac{ik}{2(z-z')}\rho^2}\nonumber \\
    \times \left[ \mathcal{I}_0\left(\frac{R_c^2(z') k^2}{8(z-z')^2} \rho^2\right) - \mathcal{I}_1\left(\frac{R_c^2(z') k^2}{8(z-z')^2} \rho^2\right) \right] e^{i(kz-\xi(z'))}. \nonumber
\end{eqnarray}
Here, $k=2\pi/\lambda$ is the wavenumber, $w(z')=w_0\sqrt{1+(z'/z_0)^2}$ is the beam radius, with $z_0=\pi w_0^2/\lambda$ the Rayleigh length, $\frac{1}{R_c(z')^2}=\frac{1}{w(z')^2}-\frac{ik}{2R(z')}-\frac{ik}{2(z-z')}$, with $R(z')=z'(1+(z_0/z')^2)$ the wavefront curvature, $\xi(z')=\tan^{-1}(z'/z_0)$ is the Gouy phase, and $\mathcal{I}_l$ is the $l-$th order modified Bessel function of the first kind.
Then, the intensity profile of the vortex beam can be calculated by taking the absolute value squared of the electric field,
\begin{equation}
\label{eq:IfarfromVR}
    I(\vec{r}) = \left|E_{x}\right|^2 + \left|E_{y}\right|^2 = \left|\tilde{E}(\rho,z)\right|^2.
\end{equation}
Figure \ref{fig:beam_generation}b and \ref{fig:beam_generation}c show the calculated 2D profiles at positions $z=0.05\,z_0$ and $z=0.5\,z_0$, respectively, together with 1D line-scans passing through the beam center. Here, the vortex retarder is placed at the beam waist ($z'=0$). Diffraction fringes are clearly visible in the first case, corresponding to the near-field region, but become less relevant as the plane of observation moves away, until they finally disappear. In both cases, an intensity singularity appears at the beam center, which is seen as point-like in the 2D intensity profiles, but actually corresponds to a line of zero intensity along the beam propagation direction.

With the intention of using a vortex beam to shape an atomic cloud, we now characterize its intensity profile close to the center, which is where the interaction between the vortex beam and the atoms takes place.
To this end, we calculate the lowest order Taylor expansion in $\rho$ of Eq.~\eqref{eq:IfarfromVR} around the beam center at $\rho=0$.
In Appendix \ref{appendix:VortexAnalytic}, we show that the intensity profile of the vortex beam close to the center can be approximated by a parabolic function $I\approx\frac{1}{2}\alpha\rho^2$. In the case of $z'\ll z_0$, meaning that the vortex retarder is placed well within the Rayleigh length from the waist of the incoming Gaussian beam, the radial intensity curvature $\alpha$ can be expressed analytically as (cf.~Appendix \ref{appendix:VortexAnalytic})
\begin{equation}
    \alpha = \frac{\pi}{\lambda\,(z-z')\,w_0^2} \left[1 + \frac{z^2-z'^2}{z_0^2}\right]^{-\frac{3}{2}} \,P.
    \label{eq:curvatureSimple}
\end{equation}
Note that the beam curvature scales linearly with the laser power $P$.

This result applies for vortex beams generated with a vortex retarder or a spiral phase plate, both of first order. In fact, their intensity profile is the same regardless of the different type of the central singularity, which is due to the spatially inhomogeneous polarization or phase.
Therefore, both beam types could be used to create thin line-like atomic distributions, as explained in section \ref{subsection:cloud_shaping}.
However, beams with a spatially dependent polarization offer an advantage for further beam shaping using e.g. polarization filters. This is illustrated in Fig.~\ref{fig:beam_generation}, which shows how a Gaussian beam with initial linear polarization (a) propagating through a vortex retarder, produces a vortex beam with purely radial polarization (b and c), which can be transformed into a double-lobed structure via a polarizer (d). 
Here, the light field at the two lobes has opposing polarization, which can be viewed as a phase shift of $\pi$, and features a line of zero intensity passing through the center of the beam, whose orientation depends on the orientation of the polarizer.
Again, we emphasize here that the zero-intensity line seen in the 2D profile of the double-lobed beam actually corresponds to a zero-intensity plane extending along the beam propagation direction. 
Assuming that the polarizer blocks the $\hat{y}$ component of the electric field, the final beam intensity can be obtained from Eq.~\eqref{eq:EfarfromVR2} as
\begin{equation}
\label{eq:IfarfromVRburger}
    I'(\vec{r}) = \left|E_{x}\right|^2 = \left|\tilde{E}(\rho,z)\right|^2 \, \cos^2{\varphi}.
\end{equation}
In general, the beam curvature $\alpha$ at the center of this beam is now a function of the coordinate $y$. However, in first order approximation this dependency can be neglected and the intensity profile close to the center can be described by a parabolic function $I\approx\frac{1}{2}\alpha x^2$. 
As in the case of vortex beams, this curvature scales linearly with the beam power, making double-lobed beams also suitable for shaping atomic clouds using the methods described in this paper, allowing for the creation of thin atomic sheets.

It should be noted that a Gaussian beam with a different orientation of its linear polarization than the one shown in Fig.~\ref{fig:beam_generation}a will produce a vortex beam with a different polarization distribution (e.g. purely azimuthal or with both radial and azimuthal components), but can always be converted into a double-lobed beam with a polarization filter.

\subsection{Cloud Shaping}
\label{subsection:cloud_shaping}

Having developed a theoretical understanding of vortex beams, we now describe how they can be used to shape unconfined atomic clouds. 
Following section \ref{subsection:vortex_beams}, the intensity profile at the vortex beam center can be approximated with a parabolic function of curvature $\alpha$, which scales linearly with the beam power $P$
\begin{equation}
\label{eq:vortexParabola}
    I(\rho) = \frac{1}{2} \, \alpha \, \rho^2,
    \qquad
    \alpha = \alpha_0 \, P 
\end{equation}
with $\alpha_0$ being the curvature for unit beam power and $\rho = \sqrt{x^2+y^2}$ the radial coordinate for a beam propagating along the z-axis. In general, the curvature $\alpha$ depends on the propagation distance $z$ as the beam expands (cf.~Eq.~\eqref{eq:curvatureSimple}). However, this dependence is negligible on the scale of the atomic cloud size $\sigma_0$, with $\sigma_0\ll z_0$, hence we assume the curvature to be constant. In the following, the basic principle of cloud shaping is described in a simplified picture, which offers an easily accessible understanding of the shaping mechanism and yields the same scaling as the full model, presented later on.

Let us start by assuming the existence of a critical intensity $I_c$ above which atoms effectively interact with the beam.
Then, we can define a critical distance $\rho_c$ from the beam center at which the interaction starts
\begin{equation}
    \rho_c = \sqrt{\frac{I_c}{\alpha}} \propto \frac{1}{\sqrt{P}}.
\end{equation}
The extent $2\rho_c$ of the cloud region in which the atoms do not interact with the vortex beam can thus be arbitrarily reduced (even below the optical diffraction limit) by increasing the beam power. Removing the atoms from the bright beam regions at $\rho>\rho_c$ will then result in a thin line-like atomic distribution.

Various shaping schemes can be employed. Here, we focus on the following two.
The first is the \textit{dynamic scheme} (see Sec.~\ref{subsection:dynamic_shaping_scheme}), which exploits the momentum transfer from resonantly scattered light to accelerate and effectively remove atoms from the bright beam regions of the cloud illuminated by the vortex beam. 
The second is the \textit{dark-state scheme} (see Sec.~\ref{subsection:dark_state_pumping_scheme}), which relies on optical pumping to transfer atoms in bright beam regions to a so-called dark state, which does not interact with the vortex beam.
Given the timescales and laser powers used in the experiments, for both shaping schemes we consider only the dissipative part of the atom-light interaction.

\subsection{Dynamic Shaping}
\label{subsection:dynamic_shaping_scheme}

\begin{figure}[!t]
	\centering
	\includegraphics[]{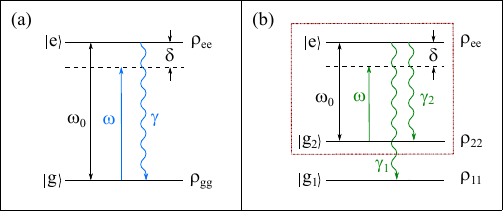}
	\caption{\textbf{Shaping schemes.} Atomic level schemes for the theoretical description of (a) the \textit{dynamic}- and (b) the \textit{dark-state} shaping methods.
	}
	\label{fig:levels_theory}
\end{figure}

For a theoretical description of the \textit{dynamic shaping scheme}, we choose an atom with a closed transition between two states, such that it can be described as a two-level systems $\ket{g}$ and $\ket{e}$, with a resonant transition frequency $\omega_0$ (cf.~Fig.~\ref{fig:levels_theory}a).

The shaping beam with frequency $\omega$ couples the two states and accelerates the atoms along its propagation direction due to the momentum transfer during successive absorption-emission cycles in the scattering process.
The force experienced by a single atom is then given by
\begin{equation}
    \vec{F}_s(\vec{r},\dot{\vec{r}}) = \hbar \vec{k} \, \Gamma(\vec{r},\dot{\vec{r}}),
\end{equation}
with $\hbar \vec{k}$ being the momentum carried by a single photon and $\Gamma(\vec{r},\dot{\vec{r}})$ the photon scattering rate.
This can be calculated as the product of the occupation probability of the excited state $\rho_{ee}$, obtained from the steady-state solution of the optical Bloch equation accounting for the Doppler effect, and its natural decay rate $\gamma$, resulting in
\begin{equation}
    \Gamma(\vec{r},\dot{\vec{r}}) = \rho_{ee}\gamma = \frac{\gamma}{2} \frac{S(\vec{r})}{1 + S(\vec{r}) + \frac{4\,\left(\delta+\vec{k} \cdot \dot{\vec{r}}\right)^2}{\gamma^2}}.
\label{eq:ScatRateDynamic}
\end{equation}
Here, $S(\vec{r})=I(\vec{r})/I_\text{sat}$ is the saturation parameter, with $I_\text{sat}$ the saturation intensity, and $\delta=\omega-\omega_0$ is the frequency detuning.
Notice that $\Gamma$ and $\vec{F}_s$ depend on both the atom position $\vec{r}$ and velocity $\dot{\vec{r}}$, due to the non-homogeneous intensity profile of the shaping beam and the influence of the Doppler shift. Taking into account the gravitational force $\vec{F}_g=m\vec{g}$ with the atomic mass $m$ and gravitational acceleration $\vec{g}$, the dynamics of the illuminated cloud is then fully described by solving the corresponding equation of motion 
\begin{equation}
\label{eq:EOMdynamic}
    \frac{\vec{F}_s\left(\vec{r},\dot{\vec{r}}\right)+\vec{F}_g}{m} = \frac{d^2\Vec{r}}{dt^2}.
\end{equation}
The complexity of Eq.~\eqref{eq:EOMdynamic} arises from the dependence of the scattering force on both the atomic position and velocity, which leads to an increasing Doppler shift and decreasing scattering rate during the time evolution. Therefore, Eq.~\eqref{eq:EOMdynamic} has to be solved for all individual atoms numerically.

\subsection{Dark-State Shaping}
\label{subsection:dark_state_pumping_scheme}

For the \textit{dark-state shaping scheme} we choose an atom with two hyperfine ground states $\ket{g_1}$ and $\ket{g_2}$ and one excited state $\ket{e}$, which allows for optical pumping between the hyperfine levels. The three-level system is illustrated in Fig.~\ref{fig:levels_theory}b, where we indicated also the time dependent populations $\rho_{ii}(t)$, $i\in\left\{1,2\right\}$ and $\rho_{ee}(t)$.
Initially, all atoms are in the upper hyperfine state, i.e. $\rho_{22}(0)=1$, $\rho_{11}(0)=\rho_{ee}(0)=0$. The vortex beam with frequency $\omega$ couples the state pair $\ket{g_2}-\ket{e}$ with a resonant transition frequency $\omega_0$, while $\ket{g_1}$ is assumed to be decoupled from the laser and therefore dark. The excited state has a total decay rate $\gamma=\gamma_1 + \gamma_2$, being the sum of the rates $\gamma_1$ and $\gamma_2$ of the individual decay channels $\ket{e}\rightarrow\ket{g_1}$ and $\ket{e}\rightarrow\ket{g_2}$, respectively.

To describe the time dynamics during the vortex illumination, we use a simple rate picture in combination with the steady-state solution of the optical Bloch equation. The latter allows to specify the time-independent relative population
\begin{equation}
\label{eq:rhoTilde}
    \tilde{\rho}=\frac{\rho_{ee}}{\rho_{22}+\rho_{ee}}=\frac{1}{2}\frac{S(\vec{r})}{1+S(\vec{r})+\frac{4\delta^2}{\gamma^2}},
\end{equation}
again with $S(\vec{r})=I(\vec{r})/I_\text{sat}$ and $\delta=\omega-\omega_0$. Note that, in contrast to Eq.~\eqref{eq:ScatRateDynamic}, we neglected here the Doppler shift, as typically only a few photons need to be scattered before the atom is pumped to the $\ket{g_1}$ dark state. With typical alkali recoil velocities $\hbar k/m$ on the order of $10\,$mm/s and resulting Doppler shifts $\hbar k^2/m$ of about $2\pi\times 10\,$kHz, this is justified, as for few scattering events the total Doppler shift stays well below the natural linewidth $\gamma$, which is typically 2-3 orders of magnitude larger. 

Using Eq.~\eqref{eq:rhoTilde}, the change in the overall population of the coupled $\ket{g_2}-\ket{e}$ system can now be described with a rate equation containing $\gamma_1$
\begin{equation}
    \frac{d}{dt}\left(\rho_{22}+\rho_{ee}\right)(t) = -\gamma_1 \rho_{ee}(t) = -\gamma_1 \tilde{\rho} \left(\rho_{22}+\rho_{ee}\right)(t).\nonumber
\end{equation}
For the initial condition $\left(\rho_{22}+\rho_{ee}\right)(0)=1$, the solution exhibits a simple exponential decay
\begin{equation}
    \left(\rho_{22}+\rho_{ee}\right)(t) = e^{-\gamma_\text{eff}t},
    \label{eq:2-level-system}
\end{equation}
where we have defined the effective decay rate of the coupled $\ket{g_2}-\ket{e}$ system to the dark ground state $\ket{g_1}$ as $\gamma_\text{eff}=\gamma_1 \tilde{\rho}$.
The time evolution for the populations of the individual states are then

\begin{equation}
\label{eq:dark_populations}
    \begin{aligned}
        \rho_{11}(t) &= 1 - e^{-\gamma_\text{eff}t}  \\
        \rho_{22}(t) &= \left(1 - \tilde{\rho} \right) e^{-\gamma_\text{eff}t}\\
        \rho_{ee}(t) &= \tilde{\rho} \, e^{-\gamma_\text{eff}t}.
    \end{aligned}
\end{equation}
Notice that Eqs. \eqref{eq:dark_populations} only represent an approximation to the real populations evolution, which is obtained by solving the set of three coupled rate equations describing the complete three-level system. However, the timescale over which the $\ket{g_2}-\ket{e}$ system reaches the steady-state is much shorter than that of the populations decay, therefore our simplified description well approximates the real time evolution.

It is now convenient to write $\gamma_\text{eff}$ as a function of the modified saturation parameter $\tilde{S}(\vec{r})=I(\vec{r})/\tilde{I}_{sat}$, where the non-resonant saturation intensity is $\tilde{I}_{sat} = I_{sat}\left( 1+ 4\delta^2/\gamma^2\right)$. Using Eq.~\eqref{eq:rhoTilde}, we obtain
\begin{equation}
\label{eq:gamma_eff}
    \gamma_\text{eff}(\vec{r}) = \gamma_1 \tilde{\rho}=\frac{\gamma_1}{2} \, \frac{\tilde{S}(\vec{r})}{1+\tilde{S}(\vec{r})}.
\end{equation}
In the case of a shaping vortex beam, the intensity profile at the beam center is approximately parabolic (cf.~Eq.~\eqref{eq:vortexParabola}), such that the saturation parameter also has a parabolic shape. Assuming that the beam propagates along the z-axis
\begin{subequations}
    \begin{equation}
      \label{eq:satParam}
        \tilde{S}(x,y) = \frac{I(x,y)}{\tilde{I}_\text{sat}} = \frac{1}{2}\,\tilde{\beta}_0\,P\,(x^2+y^2)
    \end{equation}
    \begin{equation}
    \label{eq:beta0}
        \tilde{\beta}_0 = \frac{\alpha_0}{\tilde{I}_\text{sat}}.
    \end{equation}
\end{subequations}
Combining Eqs.~\eqref{eq:gamma_eff} and \eqref{eq:satParam}, we can se that the atoms in the cloud are pumped to the $\ket{g_1}$ dark state in a spatially-dependent way, with the ones at the center of the vortex beam remaining unperturbed.

To find an analytical expression for the size of the shaped atomic cloud after some vortex beam illumination time $\tau_\text{ill}$, we assume a Gaussian density profile $\rho_0(\vec{r})$ for the initial cloud
\begin{equation}
    \rho_0(\Vec{r}) = \rho_0 \, \exp\left(-\frac{x^2}{2\sigma_{0,x}^2} - \frac{y^2}{2\sigma_{0,y}^2} - \frac{z^2}{2\sigma_{0,z}^2}\right)
\end{equation}
with characteristic widths $\sigma_{0,i}$, $i\in\left\{x,y,z\right\}$. 
We also assume the cloud expansion during the vortex beam illumination to be negligible, which is valid in the limit of $\tau_\text{ill}\ll\sigma_{0,i}\sqrt{m/k_B T}$ for a Boltzmann-Gaussian velocity distribution \cite{ketterle1999}. 
For a vortex beam centered at the cloud position, the atomic distribution after $\tau_\text{ill}$ can then be calculated as
\begin{eqnarray}
    \rho(\vec{r},\tau_\text{ill}) &=& \rho_0(\vec{r}) \, e^{-\gamma_\text{eff}\tau_\text{ill}} \\
    &\approx& \rho_0(\Vec{r}) \, \exp\left(- \frac{\gamma_1}{2} \, \tilde{S}(x,y)\, \tau_\text{ill}\right),
    \label{eq:darkRho}
\end{eqnarray}
where we have assumed $\tilde{S}\ll1$ in Eq.~\eqref{eq:gamma_eff} close to the center of the vortex beam.
Inserting $\tilde{S}(x,y)$ from Eq.~\eqref{eq:satParam} and defining
\begin{equation}
\label{eq:sigmaS}
    \sigma_s(P,\tau_\text{ill}) = \left(\frac{\gamma_1}{2}\, \tilde{\beta}_0\,\tau_\text{ill}\,P\right)^{-\frac{1}{2}},
\end{equation}
the exponential function in Eq.~\eqref{eq:darkRho} can be expressed as a Gaussian of width $\sigma_s$ in the coordinates $x$ and $y$, such that the final atomic distribution becomes
\begin{equation}
    \rho(\Vec{r},\tau_\text{ill}) \approx \rho_0 \exp\left( -\frac{x^2}{2\sigma_x^2} - \frac{y^2}{2\sigma_y^2} - \frac{z^2}{2\sigma_{0,z}^2}\right), 
\end{equation}
with
\begin{equation}
\label{eq:sigmaEff1}
    \frac{1}{\sigma_i^2} = \frac{1}{\sigma_s^2} + \frac{1}{\sigma_{0,i}^2}
    \text{,}\hspace{0.5cm}
    i\in\{x,y\}.
\end{equation}
Hence,the vortex beam shapes the density distribution of the atomic cloud in the x-y plane, while leaving it unaffected along z. Inserting Eq.~\eqref{eq:sigmaS} into Eq.~\eqref{eq:sigmaEff1} and substituting the definition of $\tilde{\beta}_0$ from Eq.~\eqref{eq:beta0}, we find the effective width $\sigma_i$ along the axis $i\in\{x,y\}$ as
\begin{equation}
\label{eq:sigmaeff}
    \sigma_i = \sigma_{0,i} \, \left(1 + \sigma_{0,i}^2\,\frac{\gamma_1}{2} \frac{\alpha_0}{I_\text{sat}\left(1+\frac{4\delta^2}{\gamma^2}\right)}\, E_\text{ill}\right)^{-\frac{1}{2}},
\end{equation}
where we defined the total energy of the vortex pulse as $E_\text{ill}=P\,\tau_\text{ill}$.
In the limit of large energies, the shaped cloud width does not depend on the initial size anymore, but only on the beam parameters
\begin{equation}
    \sigma_i \approx \sigma_s  = \left(\frac{\gamma_1}{2} \frac{\alpha_0}{I_\text{sat}\left(1+\frac{4\delta^2}{\gamma^2}\right)}\, E_\text{ill}\right)^{-\frac{1}{2}} \propto E_\text{ill}^{-\frac{1}{2}}.
\end{equation}
According to this theoretical model, the size of the shaped cloud follows a power-law dependence on the energy of the vortex pulse and can therefore be reduced arbitrarily for increasing $E_\text{ill}$.

\section{Experimental Setup}
\label{section:Experiment}

\begin{figure*}[!t]

	\centering
	\includegraphics[width=1\textwidth]{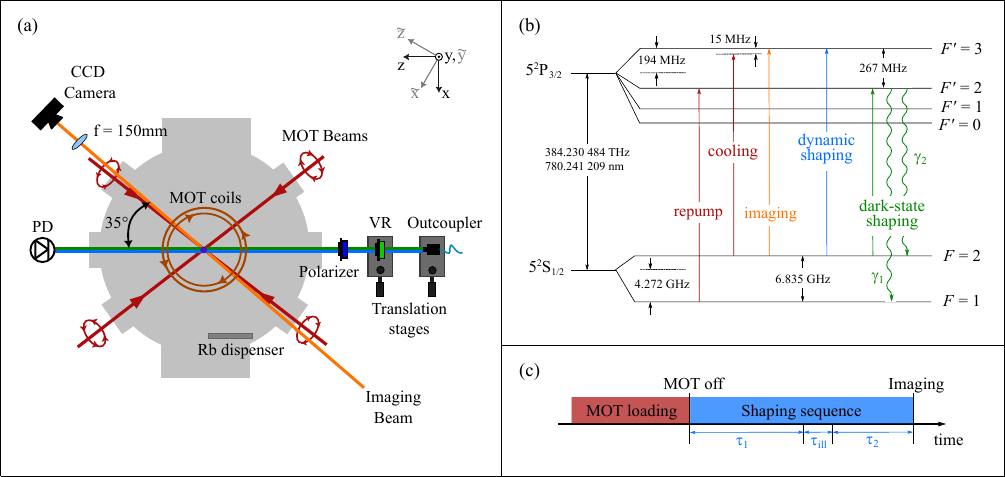}
	\caption{\textbf{Experimental realization.} (a) Experimental setup for the generation and shaping of an atomic cloud. A six-beam magneto-optical trap (MOT) is operated inside a vacuum chamber, whose lateral view-ports allow optical access with the lasers required for MOT operation, cloud imaging and shaping. The Gaussian intensity profile of the shaping beam coming out of the fiber is transformed into a vortex via the vortex retarder (VR) or into a double-lobed structure via the additional polarizer. The beam power during the shaping pulse is monitored via a photodiode (PD). Gravity is pointing in negative y-direction. (b) Level scheme of the $^{87}$Rb D2-line and transition frequencies for the lasers used in the experiment. The shaping light is tuned to different transitions, depending on the shaping method used. (c) Measurement procedure for the shaping of an atomic cloud released from a MOT. After the MOT loading, the cloud is left free to thermally expand during $\tau_1$, after which it is illuminated by the shaping light during $\tau_\text{ill}$ and then freely expands again during $\tau_2$, before being imaged.}
	\label{fig:setup} 
\end{figure*}

Our experimental setup used for the generation and shaping of atomic clouds is illustrated in Fig.~\ref{fig:setup}a. It consists of a standard six-beam magneto-optical trap (MOT) for $^{87}$Rb atoms, operated inside a vacuum chamber at a pressure of $2 \cdot 10^{-10}\,$mbar. All lasers required for MOT operation, cloud shaping and cloud imaging are guided via optical viewports into the chamber and operated near the rubidium D$_2$ line, with the specific transition frequencies given in Fig.~\ref{fig:setup}b. The cooling beam with a waist of $8\,$mm and a total power of $150\,$mW is $15\,$MHz red detuned with respect to the $\ket{5S_{1/2},F=2} \rightarrow \ket{5P_{3/2},F'=3}$ cooling transition. Atoms that are off-resonantly excited to the $\ket{5S_{1/2},F'=2}$ state and decay to the $\ket{5S_{1/2},F=1}$ ground state are pumped back to the cooling cycle via a repumping beam of $8\,$mm waist and $10\,$mW power, which is tuned resonant to the $\ket{5S_{1/2},F=1} \rightarrow \ket{5P_{3/2},F'=2}$ transition. 

For MOT operation, the cooling and repumping light are superimposed and split into three counter-propagating beam pairs, mutually overlapping at the center of the vacuum chamber. Here, a pair of Anti-Helmholtz coils creates a magnetic quadrupole field with vanishing field at the center and field gradients of $7.5$ and $15\,$G/cm in x-z and y-direction, respectively. Altogether, this creates an optical molasses, which allows for trapping and cooling atoms, loaded directly from a Rb dispenser source nearby \cite{fortagh1998}.
To reduce the temperature of the atomic cloud below the Doppler limit, we perform polarization gradient cooling during the last phase of the MOT loading. For this, the magnetic fields are turned off, except for compensation fields, and the detuning of the cooling light is increased to $\approx150\,$MHz below the atomic transition frequency. The generated cloud represents the starting point for the shaping experiments, after which the cloud is imaged via standard absorption imaging.
For this, the cloud is illuminated with an imaging beam of 5$\,$mm waist and $80\,\upmu$W power, tuned resonant to the $\ket{5S_{1/2},F=2}\rightarrow\ket{5P_{3/2},F'=3}$ transition, and the resulting beam transmission $G(\tilde{x},\tilde{y})$ is measured on a CCD camera. 
Together with two reference images (a background image $B(\tilde{x},\tilde{y})$ without atoms and a dark image $D(\tilde{x},\tilde{y})$ without light), this allows to determine the integrated atomic density $n_\text{2D}(\tilde{x},\tilde{y})$ in the plane perpendicular to the imaging axis $\tilde{z}$ \cite{ketterle1999} 
\begin{equation}
n_\text{2D}(\tilde{x},\tilde{y})\!=\!\int\!\!n(\vec{r})\,d\tilde{z}\!=\!-\frac{1}{\sigma_\text{Rb}}\!\ln\frac{G(\tilde{x},\tilde{y})-D(\tilde{x},\tilde{y})}{B(\tilde{x},\tilde{y})-D(\tilde{x},\tilde{y})},
\end{equation}
with the photon scattering cross section $\sigma_\text{Rb}$. This is a direct consequence of Lambert-Beers law, under which the beam transmission decays exponentially with the optical depth $\sigma_\text{Rb} n_\text{2D}(\tilde{x},\tilde{y})$.

The shaping beam propagates along the z-axis, forming an angle of about $35^\circ$ with the imaging beam (cf.~Fig.~\ref{fig:setup}a), and is stabilized to different transitions depending on which shaping method is used (cf.~Fig.~\ref{fig:setup}b). For the \textit{dynamic scheme}, we use the closed $\ket{5S_{1/2},F=2}\rightarrow\ket{5P_{3/2},F'=3}$ transition. For the \textit{dark-state scheme}, atoms are excited from the $\ket{5S_{1/2},F=2}$ to the $\ket{5P_{3/2},F'=2}$ state and therefore effectively pumped to the $\ket{5S_{1/2},F=1}$ ground state, which is dark to both shaping and imaging light. For the generation of the vortex beam, we employ a zero-order vortex half-wave retarder with $m=1$, which transforms the collimated Gaussian beam, with waist $w_0=1\,$mm, coming out of a polarization maintaining (PM) fiber into the desired vortex intensity profile. In the case of the \textit{dynamic shaping scheme}, for better visualization, an additional polarization filter is placed after the vortex retarder in order to create a double-lobed beam, which allows for the creation of a thin atomic sheet. 
Both the fiber outcoupler and the vortex retarder are placed on a x-y microtranslation stage to ensure that the shaping beam is centered at the atom cloud and the vortex retarder, respectively.

The measurement procedure, highlighted in Fig.~\ref{fig:setup}c, is as follows. 
First, the MOT is loaded for appropriate times up to $14.5\,$s, including polarization gradient cooling for $1.5\,$ms. The switching off of the cooling and repump beams defines the start of the shaping sequence for the unconfined cloud. The atomic cloud is first left free to expand for a time $\tau_1$, such that at the end it attains an approximately 3D Gaussian distribution. Next, the cloud is illuminated by the shaping beam, which is pulsed for a variable time $\tau_\text{ill}$. Finally, the cloud again freely evolves for a waiting time $\tau_2$ (only used for the dynamic shaping scheme), after which it is imaged.
The pulse duration $\tau_\text{ill}$ and the relative beam power are monitored by a fast photodiode, which is placed at the end of the shaping beam path outside the MOT chamber.

\section{Results}
\label{sec:results}

\subsection{Dynamic shaping}
\label{subsection:dynamic_results}

\begin{figure*}[tbp]
	\centering
	\includegraphics[]{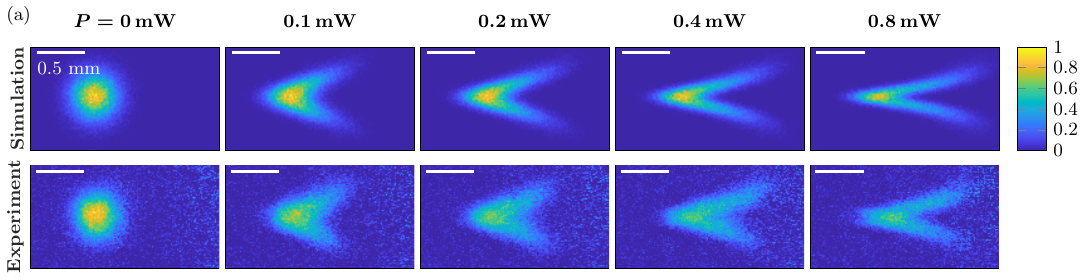}
	\par\vspace{0.5cm}
	\includegraphics[]{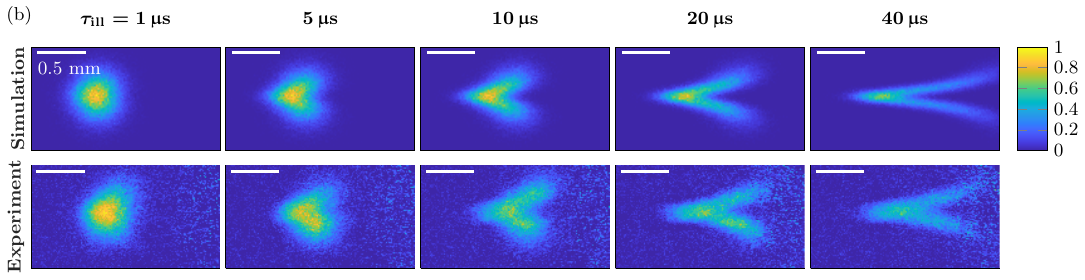}
	\par\vspace{0.5cm}
	\includegraphics[]{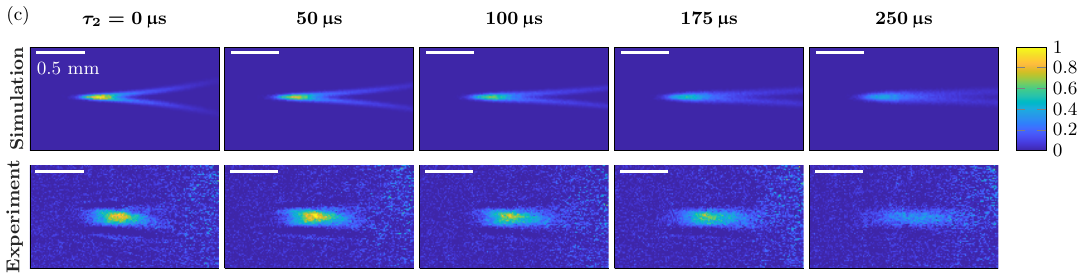}
	\caption{\textbf{Dynamic shaping with double-lobed beam.} Absorption images as simulated (upper rows) and measured (lower rows) for a series of increasing (a) shaping beam power $P$, with $\tau_1=1.2\,$ms, $\tau_\text{ill}=35\,\upmu$s, $\tau_2=565\,\upmu$s, (b) illumination time $\tau_\text{ill}$, with $\tau_1=1.2\,$ms, $\tau_2=565\,\upmu$s, $P=1.45\,$mW, and (c) free expansion time after the shaping $\tau_2$, with $\tau_1=200\,\upmu$s, $\tau_\text{ill}=300\,\upmu$s, $P=1.45\,$mW. All images are shown in the $\tilde{\text{x}}$-$\tilde{\text{y}}$ plane, with the shaping beam tuned to resonance and irradiated from the left. In the simulated images, we set $\beta_0=2.2\cdot10^{4}\, \text{mW}^{-1}\text{cm}^{-2}$ and the visibility of the atoms is adjusted depending on their final velocity to account for the Doppler shift with respect to the imaging beam. All images are normalized to the maximum atom density within the first image of each individual row, both for simulations and experiments. 
	}
	\label{fig:dynamic_images} 
\end{figure*}

For the \textit{dynamic shaping}, we present experimental results and compare them to numerical simulations introduced in Sec.~\ref{subsection:dynamic_shaping_scheme}. In particular, we investigate the dependence of the atoms dynamics on the choice of different experimental parameters, such as the beam power and the timings during the shaping sequence (cf.~Fig.~\ref{fig:setup}c), with the shaping light being tuned to resonance.
For better visualization of the cloud dynamics, we use a double-lobed beam, as shown in Fig.~\ref{fig:beam_generation}d, and rotate its intensity zero plane via the polarizer to be parallel to the x-z plane (cf.~Fig.~\ref{fig:setup}a). This means that the cloud is shaped along the vertical y-axis, with the atoms being accelerated along the propagation direction of the shaping beam (z). 
For every choice of parameters, the cloud is imaged in the $\tilde{\text{x}}$-$\tilde{\text{z}}$ plane via absorption imaging.
For the measurements presented in this section, the starting point of the shaping sequence is a cloud of $\approx5\cdot10^5$ atoms at a temperature $T\approx45\,\upmu$K, with an initial size in the vertical direction $\sigma_{0,y}\approx100\,\upmu$m.

In order to reproduce these conditions in the numerical simulations, we consider a cloud of $N$ atoms,
whose initial positions along the three axes are sampled from Gaussian distributions of widths $\sigma_{0,i}=\sigma_{0,y}$ (assuming an isotropic cloud), while the initial velocities are sampled from a three-dimensional Boltzmann distribution with temperature $T$, i.e. a Gaussian with width $\sigma_v=\sqrt{k_B T/m}$.
The shaping beam is approximated by a parabolic intensity profile $I(x,y)=\frac{1}{2}\alpha_0 P y^2$, where $\alpha_0=(1.55\pm0.02)\cdot10^{5}\,\text{cm}^{-4}$, with the error in a $95\%$ confidence interval. This value is obtained from a fit to a vertical line-scan passing through the center of the beam, imaged via a CMOS camera as it appears at the cloud position, and is assumed to be constant on the spatial scale of the cloud. In the simulations, the relevant quantity is the curvature per power of the saturation parameter $S(x,y)=I(x,y)/I_\text{sat}$, defined as $\beta_0=\alpha_0/I_\text{sat}$, which can be adapted to get a good qualitative match with the experimental pictures and therefore obtain an estimation of the saturation intensity for the driven transition.
The equation of motion \eqref{eq:EOMdynamic} is solved for the entire duration of the shaping sequence, using the ode45 solver from Matlab. In our specific case, the gravitational force is directed along the vertical y-axis, while the scattering force acts along the z-axis and is non-zero only during the illumination time $\tau_\text{ill}$.
To allow for a direct comparison of the simulated cloud with the experimental measurement, the x-y view of the resulting atomic distribution is rotated by $35^\circ$ into the imaging plane $\tilde{\text{x}}$-$\tilde{\text{y}}$ (cf.~Fig.~\ref{fig:setup}a).
Additionally, the visibility of each atom is adjusted depending on its final velocity, in order to account for the Doppler shift $\delta_D$ with respect to the imaging beam. In fact, from our simulations, we find that at the end of the shaping pulse the atoms can reach velocities along the z-axis of up to 15\,m/s, which translates to a Doppler shift of about $\delta_D=2\pi\times16\,$MHz when accounting for the $35^\circ$ angle formed by the imaging beam. This is about a factor of three larger than the natural linewidth of the imaging beam transition, which according to the Lorentzian absorption profile, reduces the atoms visibility (photon scattering rate) to  about $3\%$ of its original value. 

Figure \ref{fig:dynamic_images} shows the results of measurement series and corresponding simulations for the cloud shaping with a resonant beam for increasing (a) beam power $P$, (b) illumination time $\tau_\text{ill}$ and (c) free expansion time after shaping $\tau_2$.
In the simulations, we fix $\beta_0=2.2\cdot10^{4}\, \text{mW}^{-1}\text{cm}^{-2}$, which results in a saturation intensity of $I_\text{sat}=7.2\,\text{mW}\,\text{cm}^{-2}$.
This value is about double the expected saturation intensity of $3.3\,\text{mW}\,\text{cm}^{-2}$, which is calculated for the driven transition considering the steady-state population distribution between the magnetic sublevels (cf.~Appendix \ref{appendix:Isat}).
This discrepancy could be attributed to a non-perfect alignment between the shaping beam and the cloud in x-direction: due to the finite size of the beam, a horizontal offset of about $350\,\upmu$m is sufficient to reduce the curvature per power $\alpha_0$ by a factor 2, hence doubling the estimated saturation intensity for the chosen $\beta_0$.

For the first measurement series, the timings in the shaping sequence are set to $\tau_1=1.2\,$ms, $\tau_\text{ill}=35\,\upmu$s, $\tau_2=565\,\upmu$s and the cloud is imaged at different beam powers $P$, ranging from 0 to $0.8\,$mW. The experimental measurements and numerical simulations show very good agreement (cf.~Fig.~\ref{fig:dynamic_images}a), showing how atoms are pushed away from the upper and lower regions of the cloud in the shape of two wings.
A first aspect to be noticed is that the size of the remaining atomic layer decreases with increasing power, as expected from increasing the curvature at the center of the shaping beam. 
Furthermore, some considerations can be made regarding the shape of the lateral wings. In particular, the atomic displacement along the propagation direction of the shaping beam (z) is linearly proportional to the scattering force experienced during the shaping pulse, for which two extreme conditions can be considered.
In the limit of $I\ll I_\text{sat}$, the scattering rate, and hence the scattering force, scales linearly with the intensity (cf.~Eq.~\eqref{eq:ScatRateDynamic}), which for a double-lobed beam increases quadratically with the distance of the atoms from the beam center along the vertical y-axis. Therefore, the wings formed by the accelerated atoms should follow a square-root profile (simulations show this effect being mainly visible for powers lower than $0.1\,$mW, which can just be noticed in Fig.~\ref{fig:dynamic_images}a).
On the contrary, for $I\gg I_\text{sat}$, the force is constant independently of the atom position, ideally resulting in a constant displacement for all atoms away from the beam center. The plots in Fig.~\ref{fig:dynamic_images}a show that the wings start bending upwards for increasing beam powers, which signifies that saturation effects are becoming relevant. However, due to the power broadening, full saturation of the trajectories will only be reached for much larger powers.

A good agreement between measurements and simulations is also found for the second measurement series (cf.~Fig.~\ref{fig:dynamic_images}b), for which we still use $\tau_1=1.2\,$ms and $\tau_2=565\,\upmu$s, but fix the beam power to $P=1.45\,$mW and vary the illumination time $\tau_\text{ill}$ in the range $1-40\,\upmu$s. Here, the scattering force acts for an increasing time, resulting in an increasing velocity at the end of the shaping pulse and therefore in a larger horizontal displacement of the outer cloud wings from the original position. 
The similarity between the results in (a) and (b) attests that there is a similarity in the roles of the beam power and the illumination time in the shaping mechanism, suggesting that a more relevant quantity is the energy transferred to the atoms during the shaping pulse $E_\text{ill}=P\,\tau_\text{ill}$, at least in the limit of short illumination times.

For the third measurement set we use again $P=1.45\,$mW, while setting $\tau_1=200\,\upmu$s, $\tau_\text{ill}=300\,\upmu$s and vary $\tau_2$ in the range $0-250\,\upmu$s. Here, the long illumination time leads to an almost complete removal of the atoms from the outer regions of the cloud, while leaving the central ones in a thin line (cf.~Fig.~\ref{fig:dynamic_images}c). 
From a comparison between measurements and calculations, it can be noticed that the measured clouds are generally thicker than the simulated ones, which we attribute to the limited resolution and imperfections of our imaging system. This hypothesis is supported by the appearance of interference patterns alongside the shaped cloud.
Also, the simulations show the moving away of the outer atoms, which is completely absent in the experimental pictures. However, the trace is greatly faded even in the simulations due to the Doppler shift caused by the large velocity of the accelerated atoms, which explains the difficulty in imaging them experimentally.
Finally, both measurements and simulations show that for increasing $\tau_2$ the cloud also slightly expands along the vertical direction and that the atomic density decreases. 

The need for a finite $\tau_2$ represents one of the main disadvantages of this shaping technique compared to the \textit{dark-state scheme} presented in the next section.
In the \textit{dynamic scheme}, this waiting time is essential for the atoms to fly away from the cloud and thus for the shaping to occur, partially hindering the thinning of the remaining atomic distribution due to thermal expansion of the cloud, as discussed in Sec. \ref{section:conclusion}.
In principle, the Doppler shift experienced by the atoms during $\tau_\text{ill}$ would be sufficient to push them out of resonance for possible subsequent laser-driven manipulations of the shaped cloud. In this way, $\tau_2$ can be kept very short, if not zero, limiting the cloud expansion. However, the magnitude of the Doppler shift strongly depends on the geometry of the experimental setup.

For completeness, we mention that a shaping series with fixed power, $\tau_\text{ill}$ and $\tau_2$ and varying $\tau_1$ was also measured. In this case, the final shape of the cloud doesn't change, but the atomic density decreases for increasing $\tau_1$ due to the initial expansion of the cloud before shaping.

\subsection{Dark-state shaping}
\label{subsection:dark_state_results}

\begin{figure*}[tbp]
	\centering
	\includegraphics[]{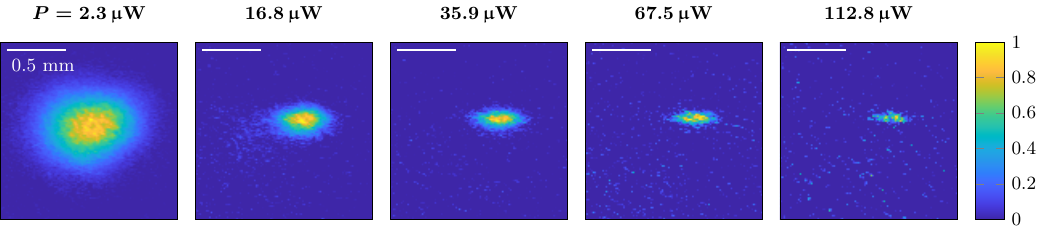}
	\caption{\textbf{Dark-state shaping with vortex beam.}
	    Absorption images for a measurement series of increasing vortex beam power $P$, with fixed timings for the shaping sequence $\tau_1=4.5\,$ms, $\tau_\text{ill}=10\,\upmu$s, $\tau_2=0$. All images result from the average of 20 pictures and are shown in the $\tilde{\text{x}}$-$\tilde{\text{y}}$ plane, normalized to the maximum atom density in each image. Between the first and the last picture, the atomic column density is reduced by more than three orders of magnitude.
	}
	\label{fig:dark_images}
\end{figure*}

\begin{figure}[tbp]
	\centering
	\includegraphics[]{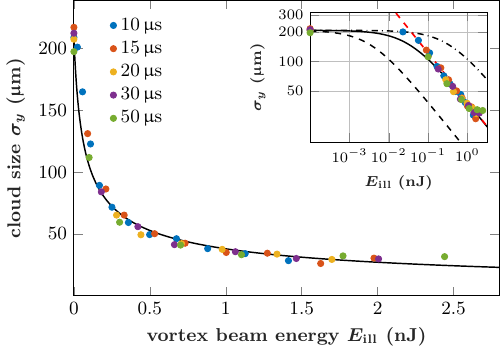}
	\caption{\textbf{Energy dependence.} 
	    Cloud size along the vertical direction $\sigma_y$ as a function of the energy of the vortex pulse $E_\text{ill}$. The colored dots represent experimental data, with the different colors corresponding to different illumination times $\tau_\text{ill}$, while the solid black line represents the fit to the common logarithm of the function in Eq.~\eqref{eq:sigmaeff}.
	    The inset shows the same plot on a log-log scale, with the dashed red line highlighting the $E_\text{ill}^{-1/2}$ dependence at high energy. The black dashed- and dashed-dotted lines show the theoretical behavior for a curvature for unit beam power of $10 \, \alpha_0$ and $\alpha_0/10$ respectively, with $\alpha_0=(1.16\pm0.02)\cdot10^5$ cm$^{-4}$ being the measured curvature for unit beam power at the center of the vortex beam (cf.~Appendix~\ref{appendix:VortexCharacterization}).
	}
	\label{fig:dark_energy}
\end{figure}

We now present experimental results and compare them to the theoretical model for the \textit{dark-state shaping method}, described in section \ref{subsection:dark_state_pumping_scheme}, for the case of a vortex beam. In particular, we aim to verify the dependence of the shaped cloud width on the energy of the vortex pulse and its frequency detuning, as described in Eq.~\eqref{eq:sigmaeff}.
For the measurements presented in this section, the starting point of the shaping sequence corresponds to a cloud of $\approx1.5\cdot10^6$ atoms at a temperature of $T\approx15\,\upmu$K, with an initial size $\sigma_{0,i}\approx200\,\upmu$m, $i\in\left\{x,y,z\right\}$.
For a simple quantitative analysis of the cloud images, we focus on the width along the vertical $\tilde{\text{y}}$-axis (corresponding to y), since the imaged horizontal atomic distribution (along $\tilde{\text{x}}$) results from a combination of the x- and z-axis due to the angle between the shaping and the imaging beam (cf.~Fig.~\ref{fig:setup}a).
To extract the cloud width, 20 absorption images are taken and averaged for every choice of parameters used for the shaping. The resulting picture is integrated along the horizontal $\tilde{\text{x}}$-axis and the integrated signal is fit with a Gaussian function, from which we deduce the vertical width $\sigma_y$.

To investigate the energy dependence, the shaping beam is tuned to resonance with the $\ket{5S_{1/2},F=2} \rightarrow \ket{5P_{3/2},F'=2}$ transition and we set $\tau_1=4.5\,$ms and $\tau_2=0$. We perform measurements at different illumination times $\tau_\text{ill}$, in the range $10-50\,\upmu$s, for which we vary the beam power $P$, ranging from 0 up to about $140\,\upmu$W. 
Figure \ref{fig:dark_images} displays some examples of averaged absorption images from the power series with $\tau_\text{ill}=10\,\upmu$s, showing the reduction of the cloud size in the vertical direction with increasing beam power.
The different measurement series can be combined in a single data set in which the varying parameter becomes the illumination energy $E_\text{ill}=P\,\tau_\text{ill}$, with values up to a maximum of about $2.4\,$nJ (for $\tau_\text{ill}=50\,\upmu$s and $P=49\,\upmu$W). The extracted widths $\sigma_y$ as a function of $E_\text{ill}$ are displayed in Fig.~\ref{fig:dark_energy}, together with the fit to the function in Eq.~\eqref{eq:sigmaeff}. 
Here, $\delta=0$, $\gamma = 2\pi\times6\,$MHz corresponds to the overall decay rate of the excited $\ket{5P_{3/2},F'=2}$ state, with $\gamma_1 = \gamma_2 = \gamma/2$, and the initial cloud size is fixed to the mean value of the extracted widths from the $P=0$ measurements, resulting in $\sigma_{0,y}=(209\pm3)\,\upmu$m.
Then, the only fit parameter is the curvature per power of the saturation parameter at the beam center $\beta_0 = \alpha_0/I_\text{sat}=(7.0\pm0.6)\cdot10^3$ mW$^{-1}$cm$^{-2}$, where the error indicates a $95\%$ confidence interval.
From characterization measurements of the vortex beam (cf.~Appendix~\ref{appendix:VortexCharacterization}), the vortex intensity curvature per power at the cloud position is estimated to be $\alpha_0=(1.16\pm0.02)\cdot10^5$ cm$^{-4}$, yielding $I_\text{sat}=(16.6\pm1.2)\,\text{mW}\,\text{cm}^{-2}$.
This is well within the interval of possible values for the saturation intensity for the $\ket{5S_{1/2},F=2}\rightarrow\ket{5P_{3/2},F'=2}$ transition, which can range between $10-20\,\text{mW}\,\text{cm}^{-2}$ depending on the initial population distribution among the magnetic sub-levels (cf.~Appendix~\ref{appendix:Isat}).
The good agreement between the experimental data and the fitted theoretical curve supports our simple theoretical model for the \textit{dark-state scheme}. Because of the radial symmetry of the vortex beam, the same behavior is expected for $\sigma_x$, while the extension of the cloud along the propagation direction z remains unaffected. Therefore, the initial atomic distribution is shaped into a cigar-like cloud, whose radial size can be decreased by increasing the shaping beam power or the illumination time.

\begin{figure}[tbp]
	\centering
	\includegraphics[]{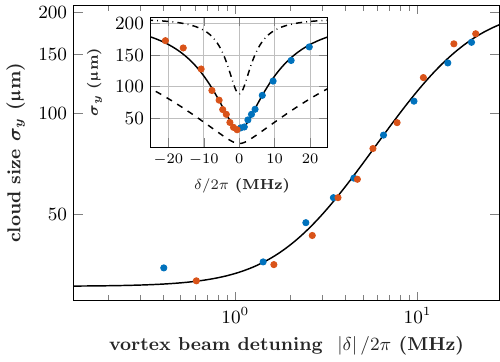}
	\caption{\textbf{Detuning dependence.} 
	    Cloud size along the vertical direction $\sigma_y$ as a function of the absolute value of the detuning of the vortex pulse $\left|\delta\right|$ (centered at zero using the fit parameter $\delta_0$), shown on a log-log scale. The colored dots represent experimental data at $E_\text{ill}=1.3\,$nJ, with the colors red and blue corresponding to red- and blue detuning respectively, while the solid black line represents the fit to the function in Eq.~\eqref{eq:sigma_detuning}.
	    The unfolded plot is shown in the inset on a lin-lin scale. The black dashed- and dashed-dotted lines show the theoretical behavior for illumination energies $10 \,E_\text{ill}$ and $E_\text{ill}/10$, respectively.
	}
	\label{fig:dark_detuning}
\end{figure}

Our model can be be further verified by investigating the dependence of the shaped cloud width on the frequency detuning of the vortex beam. In this case, we again set $\tau_1=4.5\,$ms, $\tau_2=0$ and illuminate the cloud for $\tau_\text{ill}=10\,\upmu$s with a beam power of approximately $130\,\upmu$W, corresponding to $E_\text{ill}=1.3\,$nJ. The detuning $\delta$ is varied in the range between $\pm2\pi\times 20\,$MHz, with $\delta<0$ and $\delta>0$ indicating red- and blue detunings, respectively.
The results are shown in Fig.~\ref{fig:dark_detuning}. Here, the fit function is 
\begin{equation}
    \sigma_y = \frac{\sigma_{0,y}}{ \sqrt{1 + \sigma_{0,y}^2\,\frac{\gamma_1}{2} \frac{\beta_0}{1+\left(\frac{2c\,(\delta-\delta_0)}{\gamma}\right)^2}\, E_\text{ill}}},
\label{eq:sigma_detuning}
\end{equation}
where the fit parameters $c$ and $\delta_0$ are introduced to take into account possible deviations of the frequency detuning with respect to the set value.
The unperturbed cloud size $\sigma_{0,y}$ is fixed as in the previous case and $\gamma_1=2\pi\times3\,$MHz corresponds to the rate of the single decay channel from $\ket{5P_{3/2},F'=2}$ to the dark $\ket{5S_{1/2},F=1}$ state. We obtain $c=1.4\pm0.3$, $\delta_0=2\pi\times(0.2\pm1.5)\,$MHz and $\beta_0=(9\pm2)\cdot10^3\,$mW$^{-1}$cm$^{-2}$, corresponding to a saturation intensity of $I_\text{sat} = (13\pm3)\,\text{mW}\,\text{cm}^{-2}$.
Again, this value lies within the expected interval and the good agreement between the experimental data and the fitted curve further validates our theoretical model.
Notice that the data show no difference between positive and negative detunings, meaning that only the absolute value is relevant, as predicted by the theory. Increasing the detuning has the only effect of reducing the scattering rate of the shaping beam photons on the atoms in the cloud and therefore of increasing the final cloud size.
For the creation of narrow line-shaped atomic clouds, a resonant shaping beam is thus a more convenient choice, since it allows to generate thinner distributions at a much lower beam power.

\section{Discussion and Conclusion}
\label{section:conclusion}

In this work, we have presented a method for shaping an unconfined cold atomic cloud by a vortex laser beam having a polarization singularity at the center.
We provided an analytic description for the beam electric field produced by a vortex retarder, observing that the intensity close to the center can be approximated by a parabola whose curvature scales linearly with the beam power.  
Experimentally, we demonstrated two different shaping schemes for a free expanding atomic cloud generated from a MOT of $^{87}$Rb atoms: the \textit{dynamic scheme}, in which atoms in bright beam regions are accelerated and effectively removed from the cloud, and the \textit{dark-state scheme}, in which atoms are pumped to a state not interacting with the vortex light.
In the first case, we showed a qualitative comparison of absorption imaging pictures of the shaped cloud with numerical simulations for different parameters, finding good agreement.
For the second method, we developed a simple theoretical model to quantitatively describe the thickness of the shaped cloud as a function of experimental parameters, such as the energy in the shaping pulse and the frequency detuning, which fits well our experimental data. 

The \textit{dark-state scheme} can be used to generate thin line- or sheet-like atomic clouds in the case of state-selective measurements and offers significant advantages with respect to the \textit{dynamic scheme}. First, the optical pumping to the ground dark state occurs on a much faster timescale and with a much lower beam power compared to the dynamic removal of the atoms, such that the illumination time $\tau_\text{ill}$ and $P$ can in general be reduced. Also, this method requires no additional free evolution time $\tau_2$ after the illumination pulse, preventing the atomic cloud from expanding again and therefore increasing its final size. Finally, differently from the \textit{dynamic scheme}, in this case the atoms are not physically removed from the cloud, making it possible to restore the original atomic distribution by pumping them back to the bright state, for example by using the MOT repump beam.
In view of experiments involving typical cycles of cloud shaping, manipulation of the remaining atoms and detection, high cycle rates are preferable. With a few milliseconds between individual cycles to reload the MOT from the dark state and to compensate for eventual losses, the \textit{dark-state scheme} would then allow for repetition rates of few hundreds cycles per second. 

In our work, both shaping schemes were demonstrated to enable the creation of thin atomic density distributions, for which the fundamental limitation is determined by the atomic thermal motion during the illumination time $\tau_\text{ill}$ and the waiting time $\tau_2$. For a cloud temperature of $15\,\upmu$K, corresponding to an atomic one-dimensional thermal velocity of $\sim\!\!40$ mm/s, the final cloud size due to thermal expansion amounts to $\sim\!\!15\,\upmu$m in the case of the \textit{dynamic scheme} of Fig.~\ref{fig:dynamic_images}c (with $\tau_\text{ill}+\tau_2=350\,\upmu$s), while it decreases to $\sim\!\!400\,$nm in the case of the \textit{dark-state scheme} of Fig.~\ref{fig:dark_images} (with $\tau_\text{ill}+\tau_2=10\,\upmu$s). In principle, these values can be further reduced by decreasing the initial cloud temperature. 
Currently, the limited resolution of our imaging system, estimated to be about $27\,\upmu$m, 
prevents us from detecting clouds with a thickness below $\sigma_y\approx30\,\upmu$m, as observed in Fig.~\ref{fig:dark_energy} and \ref{fig:dark_detuning}. 
For the future, we plan to overcome this limitation by means of a high resolution ion microscope developed in our laboratory, which offers a theoretical resolution limit of $100\,$nm \cite{Stecker_2017}. In any case, already now 
the fit of our theoretical model to the experimental data predicts that a cloud size on the order of $\sigma_y\approx2.7\,\upmu$m can be created with a beam energy of $E_\text{ill}\approx200\,$nJ. Considering an illumination time $\tau_\text{ill}=10\,\upmu$s, which is the shortest used in our experiment, this translates into a required power of only $20\,$mW, which is easily achievable in diode laser setups.
It is instructive to compare the achievable value for the cloud size with the one that could be reached in a diffraction-limited system by simply focusing a beam to create a dipole trap. With the focusing lens placed outside a typical vacuum chamber with a diameter of several tens of centimetres and with numerical apertures limited by CF40 viewports, the waist of a dipole trap would typically be around $20\,\upmu$m. This is well above the achievable cloud size for the \textit{dark-state scheme}. Furthermore, if we take into account the fundamental limits of our method, the achievable thicknesses would fall below this "diffraction limit" by another two orders of magnitude.

In contrast to dipole traps, which require the use of strong laser beams and optics with very large numerical aperture placed close to the atoms, our setup is very simple and the method requires very low power, making the generation of micron-scaled, unconfined clouds easily accessible. This represents a convenient experimental platform for studies with thin, unconfined atomic clouds, motivating us to use this technique for the investigation of correlation build-up through Rydberg crystallization in low-dimensional atomic ensembles. While this phenomenon has already been studied theoretically and demonstrated experimentally in optical lattices \cite{Pohl,Schauß}, it has never been observed in unconfined systems of atoms with positional disorder, i.e. with atoms not being pinned to lattice sites and free to move within the ensemble.
The realization of quasi-1D systems, with the radial extent of the cloud smaller than typical blockade distances of $\sim\!\!10\,\upmu$m, would allow to realize effective long-range interactions \cite{Weimer} between far-distant Rydberg atoms and investigate Lieb-Robinson bounds for quantum information transfer under the influence of different boundary conditions.

\section*{Acknowledgements}
The authors would like to thank Philip Osterholz for helpful discussions on the experiment. This work was supported by the Deutsche Forschungsgemeinschaft through FOR 5413.

\appendix

\section{Vortex beam - analytical treatment}
\label{appendix:VortexAnalytic}

Here, we present both a numerical and analytical treatment for the description of a vortex beam with a polarization singularity produced by a vortex retarder.
We start by considering a Gaussian TEM$_{00}$ beam propagating along the z-axis (cf.~Fig.~\ref{fig:beam_generation}), with the beam waist at position $z=0$, and a vortex retarder with negligible thickness placed in the z$'$-plane, orthogonal to the beam propagation direction. Without loss of generality, we assume the beam to be linearly polarized along the x-axis and the electric field being described by a two-component Jones vector $\vec{E}(\vec{r})=\left(E_x(\vec{r}),E_y(\vec{r})\right)$ perpendicular to the propagation axis \cite{jones1941new}. In cylindrical coordinates $\vec{r}\rightarrow (\rho,\varphi,z)$ and with the input beam polarization $\vec{p}_\text{in}=(1,0)$, the electric field in front of the retarder can then be written as
\begin{equation}
    \Vec{E}_\text{in}(\vec{r}\,') = E(\vec{r}\,')\,\vec{p}_\text{in}
\end{equation}
with
\begin{equation}
    \label{eq:Egauss}
        E(\vec{r}\,') = E_0 \frac{w_0}{w(z')} e^{-\frac{\rho'^2}{w(z')^2}} e^{i\left(kz'+\frac{k \rho'^2}{2R(z')} - \xi(z')\right)}.
\end{equation}
Here, $k=2\pi/\lambda$ is the wavenumber, $w_0$ the beam waist, $z_0=\pi w_0^2/\lambda$ the Rayleigh length, $w(z')=w_0 \sqrt{1 + (z'/z_0)^2}$ the beam radius, $R(z')= z'\, (1 + (z_0/z')^2)$ the wavefront curvature and $\xi(z')=\text{tan}^{-1}(z'/z_0)$ the Gouy phase.
The action of a vortex retarder is comparable to that of a half-wave plate for which the angle $\theta$ between the fast axis and the horizontal x-axis is a function of the azimuthal angle $\varphi$. For a vortex retarder of order $m\in\mathbb{N}$, we have
\begin{equation}
    \theta(\varphi) = \frac{1}{2}m\varphi,
\end{equation}
with the corresponding Jones-Matrix of a half-wave plate \cite{hecht2023optik}
\begin{equation}
    J_{\lambda/2}\left(\theta\right) = \begin{pmatrix} \cos{2\theta} & \sin{2\theta} \\ \sin{2\theta} & -\cos{2\theta} \end{pmatrix}.
\end{equation}
The electric field immediately after the plate can then be expressed as
\begin{equation}
\label{eq:EafterVR}
    \vec{E}_\text{out}(\vec{r}\,') = J_{\lambda/2}\left(\theta(\varphi')\right)\,\vec{E}_\text{in}(\vec{r}\,') =  E(\vec{r}\,')\,\vec{p}_\text{out}(\varphi')
\end{equation}
with $E(\vec{r}\,')$ from Eq.~\eqref{eq:Egauss} and
\begin{equation}
    \vec{p}_\text{out}(\varphi') = J_{\lambda/2}\left(\theta(\varphi')\right) \, \vec{p}_\text{in} = \begin{pmatrix} \cos{m\varphi'} \\ \sin{m\varphi'} \end{pmatrix}.
\end{equation}

The intensity profile of the vortex beam along its propagation direction can be calculated numerically making use of Huygens principle \cite{born2020}, according to which the wavefront of the propagating beam is the result of the interference of secondary spherical wavelets originating from the initial wavefront.
The field at position $\vec{r}$ can be expressed as
\begin{equation}
    \label{eq:HuygensIntegral}
    \vec{E}_\text{out}(\vec{r}) = -\frac{i}{\lambda}\iint\, \vec{E}_\text{out}(\vec{r}\,')\, S(\vec{r}-\vec{r}\,') \, dx'\, dy',
\end{equation}
which corresponds to the two-dimensional convolution of the original field with the spherical wavelet $S(\vec{r})$
\begin{equation}
   S(\vec{r}) = \frac{e^{ikr}}{r}\approx\frac{1}{z}\,e^{ikz}\,e^{ik\frac{\rho^2}{2z}},
\end{equation}
with $r=\left|\vec{r}\right|=\sqrt{\rho^2+z^2}$ and the Fresnel approximation being used in the limit $\rho\ll z$, at which the wavefronts become parabolic. 
The overall intensity of the vortex beam is then simply obtained by adding the absolute values squared of the electric field components
\begin{equation}
    I_\text{out}(\vec{r}) \,\sim\,  \left|E_{\text{out},x}(\vec{r})\right|^2 + \left|E_{\text{out},y}(\vec{r})\right|^2,
\end{equation}
with normalization to the total beam power $P$
\begin{equation}
    \iint I_\text{out}(\vec{r})\,dx\,dy = P.
\end{equation}

Next, we present an analytic description of vortex beams with a polarization singularity.
As mentioned in Sec. \ref{subsection:vortex_beams}, the intensity profile of our experimentally generated beam only resembles that of a LG$_{01}$ mode, even in the far-field.
In the near-field, diffraction effects become visible and an accurate representation can be obtained making use of the Collins-Huygens-Integral \cite{Collins}, which is equivalent to the paraxial Fresnel integral using the ray transfer matrix (also know as ABCD matrix) formalism. Again assuming that the beam propagates along the z-axis, the final electric field distribution at a position $z$ can be calculated for any known initial field at position $z'$ as a function of the ABCD matrix coefficients describing the propagation from $z'$ to $z$. In cylindrical coordinates,
\begin{eqnarray}
\label{eq:Collins}
    &&\vec{E}_\text{out}(\vec{r}) = -\frac{i}{\lambda B} e^{ik(z-z')} \int_0^\infty \rho'\,d\rho' \int_0^{2\pi} d\varphi'\, \vec{E}_\text{out}(\vec{r}\,')  \nonumber\\ 
    && \;\;\times \exp\left[\frac{ik}{2B}\left(A\rho^2+D\rho'^2 - 2\rho\rho'\cos(\varphi-\varphi')\right)\right],
\end{eqnarray}
where the propagation matrix for free space evolution is
\begin{equation}
\label{eq:ABCD}
    \begin{pmatrix} A & B \\ C & D \end{pmatrix} = 
    \begin{pmatrix} 1 & z - z' \\ 0 & 1 \end{pmatrix}.
\end{equation}
Then, Eq.~\eqref{eq:Collins} follows directly from Eq.~\eqref{eq:HuygensIntegral} for $z>z'$ and using the Fresnel approximation for the spherical wavelet, with $\left|\vec{r}-\vec{r}\,'\right|$ expressed in cylindrical coordinates
\begin{equation}
\left|\vec{r}-\vec{r}\,'\right|\approx(z-z')+\frac{\rho^2+\rho'^2-2\rho\rho'\cos{(\varphi-\varphi')}}{2(z-z')}.
\end{equation}
In the following, we will adapt the calculations for the vortex field after a spiral phase plate presented in \cite{Mawardi} to the case of a vortex retarder. Inserting Eq.~\eqref{eq:EafterVR} in Eq.~\eqref{eq:Collins}, we can write the electric field at any position after the vortex retarder as
\begin{equation}
\begin{split}
    \vec{E}&_\text{out}(\vec{r}) = -\frac{i}{\lambda B} E_{00}(\rho,z') \, e^{ikz} \int_0^\infty \int_0^{2\pi}  
    \begin{pmatrix} \cos{m\varphi'} \\ \sin{m\varphi'} \end{pmatrix}  \\
    &\times \exp\left[-\frac{\rho'^2}{R_c(z')^2}\right]
    \exp\left[-i\frac{\rho'}{\rho_c}\cos(\varphi-\varphi')\right] \,\rho'\,
   \, d\rho' d\varphi',\nonumber
\end{split}
\end{equation}
where we defined
\begin{subequations}
    \begin{equation}
        E_{00}(\rho,z') = E_0 \frac{w_0}{w(z')} e^{\frac{ikD\rho^2}{2B}} e^{-i\xi(z')}
    \end{equation}
    \begin{equation}
        \frac{1}{R_c(z')^2} = \frac{1}{w(z')^2} - \frac{ik}{2R(z')} - \frac{iAk}{2B}
    \end{equation}
    \begin{equation}
        \frac{1}{\rho_c} = \frac{k\rho}{B}.
    \end{equation}
\end{subequations}
The integral over the azimuthal angle $\varphi'$ can be solved after substituting $\phi=\varphi-\varphi'$ and using
\begin{equation}
    \int_{\varphi-2\pi}^{\varphi} e^{-i(\kappa\cos\phi + m\phi)} \, d\phi = 2\pi (-i)^{\left|m\right|} J_{\left|m\right|}\left(\kappa\right),
\end{equation}
with $\kappa=\rho'/\rho_c$ and $J_m$ being the $m-$th order Bessel function of the first kind. This allows to write
\begin{eqnarray}
    &&\vec{E}_\text{out}(\vec{r}) = \frac{(-i)^{\left|m\right|+1}}{k B} E_{00}(\rho,z') \, e^{ikz} \begin{pmatrix} \cos{m\varphi} \\ \sin{m\varphi} \end{pmatrix}\nonumber\\
    &&\times\int_0^\infty \exp\left[-\frac{\rho'^2}{R_c(z')^2}\right] \, J_{\left|m\right|}\left(\frac{\rho'}{\rho_c}\right)
     \,\rho'\, d\rho',
\end{eqnarray}
where it can be noticed that the propagation does not affect the beam polarization $\vec{p}_\text{out}=(\cos(m\varphi'),\sin(m\varphi'))$.
This integral can be expressed in terms of a combination of $l-$th order modified Bessel functions of the first kind $\mathcal{I}_l$ making use of the formula \cite{gradshteyn} 
\begin{equation}
    \int_0^\infty\!\!\!\! x e^{-\epsilon x^2}\!J_{\left|m\right|}(\beta x) dx =\frac{\sqrt{2\pi\eta}}{4\epsilon} e^{-\eta} 
    \left[ \mathcal{I}_{\frac{\left|m\right|-1}{2}}\!(\eta)\!-\!\mathcal{I}_{\frac{\left|m\right|+1}{2}}\!(\eta) \right]\nonumber,
\end{equation}
with $\eta=\beta^2/8\epsilon$. Identifying $\epsilon=1/R_c^2$ and $\beta=1/\rho_c$ leads to an analytic expression for the electric field of a vortex beam at a position $z$ which is generated by a $m$-th order vortex retarder placed at $z'$:
\begin{eqnarray}
\label{eq:EfarfromVR}
    \Vec{E}_\text{out}(\vec{r}) = \frac{2\pi^{3/2}(-i)^{\left|m\right|+1}
    E_{00}(\rho,z')R_c^3(z')}{8\rho_c\lambda B}  \, e^{ikz} \nonumber\\
     \times e^{-\frac{R_c^2}{8\rho_c^2}}
    \left[ \mathcal{I}_{\frac{\left|m\right|-1}{2}}\!\left(\frac{R_c^2}{8\rho_c^2}\right)\!-\!\mathcal{I}_{\frac{\left|m\right|+1}{2}}\!\left(\frac{R_c^2}{8\rho_c^2}\right)\right]\vec{p}_{out}(\varphi).
\end{eqnarray}

We focus on the case of a vortex beam with $m=1$, as the one we use in our experiments.
Notice that the field dependence on the azimuthal angle $\varphi$ is only contained in the polarization vector, allowing to rewrite Eq.~\eqref{eq:EfarfromVR} as
\begin{equation}
\label{eqAPP:EfarfromVR2}
    \Vec{E}_\text{out}(\vec{r}) = \tilde{E}_\text{out}(\rho,z)  \, \begin{pmatrix} \cos{\varphi} \\ \sin{\varphi} \end{pmatrix},
\end{equation}
with
\begin{eqnarray}
    \tilde{E}_\text{out}(\rho,z) = - \sqrt{\pi} E_0 \frac{w_0}{w(z')} \frac{R_c^3(z') k^2}{8B^2} \, \rho\, e^{-\frac{R_c^2(z') k^2}{8B^2} \rho^2} e^{\frac{ikD}{2B}\rho^2}\nonumber \\
    \times \left[ \mathcal{I}_0\left(\frac{R_c^2(z') k^2}{8B^2} \rho^2\right) - \mathcal{I}_1\left(\frac{R_c^2(z') k^2}{8B^2} \rho^2\right) \right] e^{i(kz-\xi(z'))}. \nonumber
\end{eqnarray}
Then, the intensity profile of the vortex beam can be calculated by taking the absolute value squared of the electric field,
\begin{equation}
\label{eqApp:IfarfromVR}
    I_\text{out}(\vec{r}) = \left|E_{\text{out},x}\right|^2 + \left|E_{\text{out},y}\right|^2 = \left|\tilde{E}_\text{out}(\rho,z)\right|^2.
\end{equation}

With the intention of using a vortex beam to shape an atomic cloud, we now characterize its intensity profile close to the center, which is where the interaction between the vortex beam and the atoms takes place.
To this end, we calculate the lowest order Taylor expansion in $\rho$ of Eq.~\eqref{eqApp:IfarfromVR} around the beam center at $\rho=0$.
Knowing that $\mathcal{I}_0(0) = 1$, $\mathcal{I}_1(0) = 0$ and making use of the following expressions for the derivatives of the modified Bessel functions \cite{olver10}
\begin{subequations}
    \begin{equation}
        \frac{\partial}{\partial r}\mathcal{I}_0(r) = \mathcal{I}_1(r)
    \end{equation}
    \begin{equation}
        \frac{\partial}{\partial r}\mathcal{I}_m(r) = \frac{1}{2} 
        \left[\mathcal{I}_{m-1}(r) + \mathcal{I}_{m+1}(r)\right],
    \end{equation}
\end{subequations}
one can show that $I_\text{out}(\rho=0)= 0, \, \left.\frac{\partial I_\text{out}}{\partial \rho}\right|_{\rho=0}  = 0$, while
\begin{equation}
\begin{split}
    \left.\frac{\partial^2 I_\text{out}}{\partial \rho^2}\right|_{\rho=0}  &= 
    2\left.\frac{\partial\tilde{E}}{\partial \rho}\left(\frac{\partial\tilde{E}}{\partial \rho}\right)^*\right|_{\rho=0} \\
    &= 2 \pi \left| E_0 \frac{w_0}{w(z')} \, \frac{R_c^3(z') k^2}{8B^2} \right|^2.
\end{split}
\end{equation}
We can then approximate the intensity profile of the $m=1$ vortex beam close to the center with a parabolic function $I\approx\frac{1}{2}\alpha \rho^2$ and find an analytic expression for the radial intensity curvature $\alpha$
\begin{eqnarray}
\label{eq:alpha}
    &&\alpha = \left.\frac{\partial^2 I_\text{out}}{\partial \rho^2}\right|_{\rho=0} = \frac{\pi P}{\lambda\,(z-z')\,w(z')^2} \nonumber\\
    &&\times\!\left(\! 
    \left[\!1\!+\!\frac{z\!-\!z'}{z_0}\!\left(\frac{w_0}{w(z')}\right)^2\!\frac{z'}{z_0}\right]^2\!+\! \left[\!\frac{z\!-\!z'}{z_0}\!\left(\frac{w_0}{w(z')}\right)^2\right]^2
    \right)^{-\frac{3}{2}}\!\!\!\!,\nonumber
\end{eqnarray}
where we used the ABCD matrix for free-space propagation from $z'$ to $z$ in Eq.~\eqref{eq:ABCD} and $\left|E_0\right|^2= 2P/\pi w_0^2$ for the Gaussian input beam at the vortex retarder, with $P$ being the beam power.  
This expression is valid for all $z$ along the beam propagation and the most important thing to notice is that the beam curvature scales linearly with the total power.
In the case $z'\ll z_0$, meaning that the vortex retarder is placed well within the Rayleigh length from the waist of the incoming Gaussian beam, a simple expression for $\alpha$ can be obtained
\begin{equation}
    \alpha = \frac{\pi}{\lambda\,(z-z')\,w_0^2} \left[1 + \frac{z^2-z'^2}{z_0^2}\right]^{-\frac{3}{2}} \,P.
\label{eqApp:curvatureSimple}
\end{equation}

\section{Vortex beam - experimental characterization}
\label{appendix:VortexCharacterization}

\begin{figure}[tbp]
	\centering
	\includegraphics[]{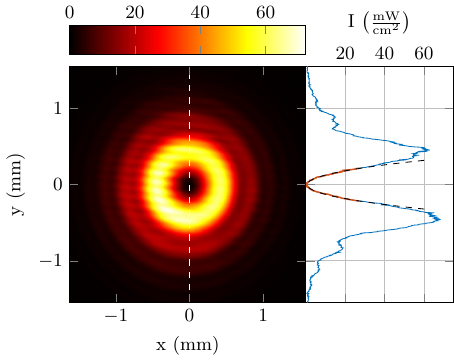}
	\caption{\textbf{Vortex beam.} Intensity profile of the vortex beam as imaged with a CMOS camera at the MOT position. The 2D profile is shown together with a 1D line-scan through the beam center ($x=0$), with the line-scan position marked in the 2D profile (white dashed line). The orange line in the 1D profile identifies the data used to fit the parabola at the beam center (black dashed line), whose curvature provides an estimation for the parameter $\alpha_0$ from Eq.~\eqref{eq:vortexParabola}. The intensity is normalized such that the integrated 2D profile leads to a total power of $1\,$mW.
	}
	\label{fig:vortex}
\end{figure}

Here, we present a characterization measurement of the vortex beam used for the dark-state shaping presented in Sec.~\ref{subsection:dark_state_results}, aiming to estimate the parameter $\alpha_0$ in Eq.~\eqref{eq:vortexParabola}. Therefore, the shaping beam is deflected in front of the chamber and imaged via a CMOS camera, placed at the same distance to the vortex retarder as the MOT.   
Thus, the beam profile resembles the one at the MOT position. To correct for the black level of the camera, we take a reference image without light, which is subtracted from the image with light in order to obtain the final picture.
Figure \ref{fig:vortex} shows the resulting 2D profile, where the pixel illumination has been normalized such that the integrated signal provides a total power $P=1\,$mW.
In order to determine the beam curvature in the parabolic approximation (cf.~Sec.~\ref{subsection:vortex_beams}), we consider the 1D intensity profile along a vertical line-scan passing through the beam center ($x=0$), i.e. along the same direction as we observe the corresponding cloud shaping in the dark-state measurements. 
A parabolic function $I=\frac{1}{2}\alpha_0 P y^2$ is fitted to the 1D profile near the vortex center, resulting in $\alpha_0 = (1.16\pm0.02)\cdot10^5\,\text{cm}^{-4}$.

\section{Saturation intensity}
\label{appendix:Isat}

Here, we estimate the saturation intensity for the $\ket{5S_{1/2},F=2}\rightarrow\ket{5P_{3/2},F'=3}$ transition, driven in the \textit{dynamic scheme}, and the interval of possible saturation intensities for the $\ket{5S_{1/2},F=2}\rightarrow\ket{5P_{3/2},F'=2}$ transition, driven in the \textit{dark-state scheme} to pump the atoms to the $\ket{5S_{1/2},F=1}$ state.

\begin{figure}[tbp]
	\centering
	\includegraphics[]{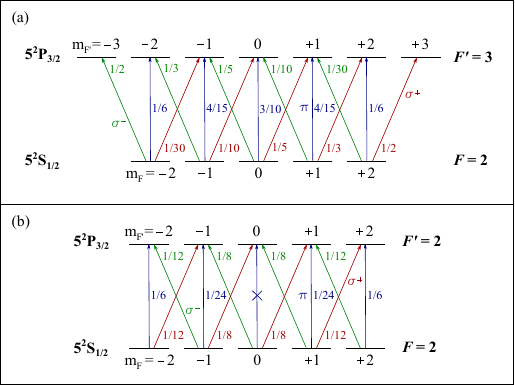}
	\caption{Level Scheme for the (a) $^{87}$Rb $\ket{5^{2}S_{1/2},F=2}\rightarrow\ket{5^{2}P_{3/2},F'=3}$ and (b) $\ket{5^{2}S_{1/2},F=2}\rightarrow\ket{5^{2}P_{3/2},F'=2}$ hyperfine transitions. Magnetic sub-levels are denoted by their respective $m_F$, $m_{F'}$ quantum numbers. Possible $\sigma^{+}$, $\sigma^{-}$ and $\pi$ transitions are shown in red, green and blue, respectively, together with their relative excitation probabilities expressed as fractions of $P_{\text{D}_2} = \left|\langle J=1/2\|er\|J'=3/2\rangle\right|^2$ \cite{Steck}. All other excitation probabilities are zero.
	}
	\label{fig:Isat} 
\end{figure}

In general, the saturation intensity is inversely proportional to the transition strength, which depends on the coupling between the light polarization and the atomic dipole moment.
As in the experiment, we consider a vortex beam with purely radial polarization (cf.~Fig.~\ref{fig:beam_generation}c) propagating along the $z$ axis. With the MOT magnetic field being turned off during the shaping pulse, we choose the quantization axis to point along the shaping beam direction. At the position of each individual atom, the linear beam polarization is then oriented orthogonal to the quantization axis, driving $\sigma^+$ and $\sigma^-$ transitions at half the beam power each. 
Figure \ref{fig:Isat} shows a schematic of the atomic sub-levels for the given ground and excited states, together with the possible transitions driven by the shaping light and the relative excitation probabilities, which are expressed as fractions of the absolute square of the reduced transition dipole matrix element for the D$_2$ line ($P_{\text{D}_2} = \left|\langle J=1/2\|er\|J'=3/2\rangle\right|^2$ \cite{Steck}).  

The coupling strength for an $\ket{F}\rightarrow\ket{F'}$ transition depends in general on the initial population distribution among the magnetic sub-levels, such that we can express the average coupling between the levels in terms of an average transition dipole moment, defined as
\begin{eqnarray}
\label{eq:average_dipole}
    \left|\Bar{d}\right|_i^2 &=& \left|\bra{F} d \ket{F'}\right|_i^2 \\
    &=& \!\!\!\!\sum_{m_F,m_{F'},q} \!\!\!\!
    \left|\bra{F\,m_F} e r_q \ket{F'\,m_{F'}}\right|^2 \, P_{m_F}^i \, P_q.
\end{eqnarray}
Here, the first term in the summation  
is the excitation probability for the $\ket{F,m_F}\rightarrow\ket{F',m_{F'}}$ transition, with $q$ referring to the angular momentum carried by the photon, ($q=\pm1$ for $\sigma^\mp$ and $q=0$ for $\pi$ transitions).
With $P_q$ we indicate the beam power distribution between the different types of transitions, which in our case are $P_1=P_{-1}=1/2$ and $P_0=0$. Finally, $P_{m_F}^i$ refers to the initial population distribution among the $m_F$ sub-levels of the ground state.

To determine the saturation intensity for a specific transition and initial distribution, it is useful to express it in terms of a reference value $I_{\text{sat},\text{ref}}$. Due to the inverse relationship between the saturation intensity and the dipole moment, the following expression holds
\begin{equation}
\label{eq:Isat}
    \frac{I_{\text{sat},i}}{I_{\text{sat},\text{ref}}}
    =\frac{\left|\Bar{d}\right|_\text{ref}^2}{\left|\Bar{d}\right|_i^2}.
\end{equation}
As a reference we consider the $\ket{F=2,m_F=2}\rightarrow\ket{F'=3,m_{F'}=3}$ cycling transition under $\sigma^+$ polarized light, for which $I_{\text{sat},\text{ref}}=1.67\,\text{mW}\,\text{cm}^{-2}$ \cite{Steck}. In this case, the transition dipole moment $\left|\Bar{d}\right|_\text{ref}^2$ is equal to the excitation probability between the given states and can be conveniently expressed as multiple of $P_{\text{D}_2}$, resulting in $\nicefrac{1}{2}\,P_{\text{D}_2}$ \cite{Steck}.

We now want estimate the saturation intensity of the $\ket{5S_{1/2},F=2}\rightarrow\ket{5P_{3/2},F'=3}$ transition, which is driven in the \textit{dynamic scheme}. In this case, the shaping process occurs over multiple successive absorption-emission cycles, such that we can consider the population distribution as the one in the steady state. Solving the corresponding rate equations results in $P_{m_F=\pm2}=0.3638$, $P_{m_F=\pm1}=0.0924$, $P_{m_F=0}=0.0876$ in our $\sigma^+$/$\sigma^-$ pumping scheme. Following Eq.~\eqref{eq:average_dipole} and summing over all $\sigma^\pm$ transitions shown in Fig.~\ref{fig:Isat}a, we obtain $I_\text{sat}=3.3\,\text{mW}\,\text{cm}^{-2}$ independent of the initial sub-level distribution.

In order to determine limiting values for the saturation intensity of the $\ket{5S_{1/2},F=2}\rightarrow\ket{5P_{3/2},F'=2}$ transition for the \textit{dark-state scheme}, we consider two conditions for the atomic spin, namely a completely unpolarized- or polarized state, and calculate the corresponding average of the squared transition dipole moments. 
In the first case, the atomic population is equally distributed among the five ground state sub-levels, hence $P_{m_F}^{i}=1/5$. 
Following Eq.~\eqref{eq:average_dipole} and summing over all $\sigma^\pm$ transitions shown in Fig.~\ref{fig:Isat}b, we then obtain $\left|\Bar{d}\right|^2=\nicefrac{1}{12}\,P_{\text{D}_2}$. From Eq.~\eqref{eq:Isat}, this results in $I_\text{sat}=10\,\text{mW}\,\text{cm}^{-2}$.
In the second case of a fully polarized state, we assume the whole atomic population to be in the $m_F=+2$ sub-level, meaning that $P_{m_F=2}=1$ is the only non-zero contribution. Again, following Eq.~\eqref{eq:average_dipole}, we find $\left|\Bar{d}\right|^2=\nicefrac{1}{24}\,P_{\text{D}_2}$, leading to $I_\text{sat}=20\,\text{mW}\,\text{cm}^{-2}$. Depending on the specific initial state population, the saturation intensity is thus expected to be between $10-20\,\text{mW}\,\text{cm}^{-2}$. While the initial state population is not easily accessible in a magneto-optical trap, it also changes during the optical pumping. In our $\sigma^+$/$\sigma^-$ pumping scheme, a steady-state population distribution is reached after sufficient absorption-emission cycles, with $P_{m_F=\pm 2}=0.2710$, $P_{m_F=\pm 1}=0.1369$ and $P_{m_F=0}=0.1843$, resulting in a saturation intensity of $11.3\,\frac{\text{mW}}{\text{cm}^2}$. However, this value cannot be taken as a reference in our case, as the atoms will be pumped to the $\ket{5^2S_{1/2},F=1}$ dark state before this distribution is reached.

\end{document}